\documentstyle[12pt,aaspp4,flushrt,psfig]{article}
\begin{document}

\title{APM 08279+5255: Keck Near- and Mid-IR High-Resolution Imaging \altaffilmark{1}} 

\author{E.\ Egami, G.\ Neugebauer, B.\ T.\ Soifer, and K.\ Matthews}
\affil{Palomar Observatory, California Institute of Technology, 320-47, Pasadena, CA 91125\\
(egami, gxn, bts, kym)@mop.caltech.edu}

\author{M.\ Ressler}
\affil{Jet Propulsion Lab, 169-506, 4800 Oak Grove Dr., Pasadena, CA 91109\\
ressler@cougar.jpl.nasa.gov}

\author{E.\ E.\ Becklin}
\affil{Department of Physics and Astronomy, UCLA, 156205 Los Angeles, CA90095\\
becklin@astro.ucla.edu}

\author{T.\ W.\ Murphy Jr.}
\affil{Palomar Observatory, California Institute of Technology, 320-47, Pasadena, CA 91125\\
tmurphy@mop.caltech.edu}

\and

\author{D.\ A.\ Dale}
\affil{IPAC, California Institute of Technology, 100-22, Pasadena, CA
91125\\
dad@ipac.caltech.edu}

\altaffiltext{1}{Based on observations obtained at the W.\ M.\ Keck
Observatory which is operated jointly by the California Institute of
Technology, the University of California, and NASA.}

\begin{abstract} 

We present Keck high-resolution near-IR (2.2 $\mu$m; FWHM $\sim$
0\farcs15) and mid-IR (12.5 $\mu$m; FWHM $\sim$ 0\farcs4) images of
APM 08279+5255, a $z=3.91$ IR-luminous BALQSO with a prodigious
apparent bolometric luminosity of $5 \times 10^{15} {\rm L}_{\odot}$, the
largest known in the universe.  The $K$-band image shows that this
system consists of three components, all of which are likely to be the
gravitationally lensed images of the same background object, and the
12.5 $\mu$m image shows a morphology consistent with such an image
configuration.  Our lens model suggests that the magnification factor
is $\sim$ 100 from the restframe UV to mid-IR, where most of the
luminosity is released.  The intrinsic bolometric luminosity and IR
luminosity of APM 08279+5255 are estimated to be 5$\times 10^{13} {\rm
L}_{\odot}$ and 1 $\times 10^{13} {\rm L}_{\odot}$, respectively.
This indicates that APM 08279+5255 is intriniscally luminous, but it
is not the most luminous object known.  As for its dust contents,
little can be determined with the currently available data due to the
uncertainties associated with the dust emissivity and the possible
effects of differential magnification.  We also suggest that the
lensing galaxy is likely to be a massive galaxy at $z \sim 3$.

\end{abstract} 
\section{INTRODUCTION} 

In the course of a survey for distant cool carbon stars in the
Galactic halo, Irwin et al.\ (1998) serendipitously discovered an
extremely luminous broad absorption line QSO at $z=3.91$, APM
08279+5255.\footnote{The redshift obtained by Irwin et al.\ (1998) was
3.87, but the CO line observation by Downes et al.\ (1999) later
showed that the redshift of this system is 3.9110, and that the
absorption lines measured by Irwin et al.\ (1998) are probably
blueshifted due to the gas outflow in this object.  To avoid
confusion, we adopt a redshift of 3.91 throughout this paper.}  Its
phenomenally large apparent luminosity is immediately clear from its
observed magnitude: the observed $R$ magnitude of 15.2 mag corresponds
to an absolute magnitude of $M_{R}=-33.2$ mag after a K correction.
Furthermore, this object was also detected at 25, 60, and 100 $\mu$m
in the IRAS Faint Source Catalog with flux densities of 0.23, 0.51,
and 0.95 Jy.  All together, the {\em apparent} bolometric luminosity
of this object reaches an unprecedented level of $5 \times 10^{15}$
L$_{\odot}$, which would make APM 08279+5255 the most luminous object
known in the universe\footnote{Throughout this paper, we adopt the
values of H$_{0} = 50$ km s$^{-1}$ Mpc$^{-1}$ and q$_{0} = 0.5$.}.

The observations to date indicate that APM 08279+5255 is likely to be
gravitationally lensed.  Irwin et al.\ (1998) showed that the $R$-band
image of this object, taken under a 0\farcs9 seeing, is slightly
elongated, and that the image likely consists of two point sources
separated by 0\farcs3--0\farcs45 with a flux ratio of 1.05--1.15,
probably being two gravitationally lensed images of the same QSO.
Subsequently, Ledoux et al.\ (1998) performed adaptive-optics
observations in the $H$ band, which achieved a spatial resolution of
0\farcs3.  Their image shows two point sources separated by
0\farcs35$\pm$0\farcs02 with a flux ratio of 1.21$\pm$0.25, in good
agreement with the values derived by Irwin et al.\ (1998).  In
addition, by using an optical integral-field spectrograph, Ledoux et
al.\ (1998) showed that the optical spectra of these two sources are
quite similar, strengthening the gravitational lensing hypothesis for
this object.

At the same time, the far-IR and submillimeter observations have shown
that APM 08279+5255 contains a large amount of gas and dust, making
this object an extreme example of a hyper/ultra-luminous IR
galaxy/QSO.  In addition to the IRAS far-IR detections, its continuum
was also detected in the submillimeter by Lewis et al (1998).  The
black-body dust temperature was estimated from the submillimeter SED
to be 220 K, which in turn results in a dust mass with no
magnification correction of $3.7 \times 10^{9}$ M$_{\odot}$
(\cite{Lewis98}).  Later, Downes et al.\ (1999) detected two CO lines
(4--3 and 9--8) by millimeter interferometry, and derived the
molecular gas temperature of $\sim 200$ K from the line ratio.  The
mass of molecular gas was calculated to be $1-6 \times 10^{9}$
M$_{\odot}$ with a magnification factor of 7--20.  From a simultaneous
millimeter continuum observation, they also derived the dust mass of
$1-7 \times 10^{7}$ M$_{\odot}$ with a magnification factor of 7--30.
The detection of the CO (9--8) line is especially important because
this is direct evidence that this object contains hot and dense
molecular gas.

APM 08279+5255 was also found to have a significant visual linear
polarization ($p>1$~\%, Hines et al.\ 1999).  Hines et al.\ 
also suggest that the broad absorption trough formerly identified as a
$z=3.07$ damped Ly-$\alpha$ absorption system by Irwin et al.\ (1998)
may be a O~{\sc iv}/Ly$\beta$ broad absorption line intrinsic to the
QSO because there is an increase of polarization in the trough.

%Because of its exceptionally large apparent brightness, the spectrum
%of APM 08279+5255 provides an excellent opportunity to study the QSO
%absorption line systems.  A Keck HIRES spectrum of APM 08279+5255 was
%analyzed by Ellison et al. (1999a;1999b), which could potentially
%provide information on the foreground lensing galaxy.

From the currently available data, APM 08279+5255 looks very much like
the hyperluminous IR QSO H1413+117 (Cloverleaf QSO) and the warm
ultraluminous IR galaxy Mrk 231, in the sense that it contains a
clearly visible QSO nucleus surrounded by a large amount of gas and
dust.  The major difference between these two objects and APM
08279+5255 seems to be the exceptionally large luminosity of the
latter, which is likely the effects of gravitational lensing.  In
these systems, the existence of a powerful QSO nucleus seems to
indicate that the QSO is the dominant luminosity source, generating
the large IR luminosity by heating dust.

Here, we present high-resolution images of APM 08279+5255 in the near-
and mid-IR taken with the Keck telescope.  Our high-resolution (FWHM
$\sim$ 0\farcs15) $K$-band image shows a third component between the
two components previously detected.  We argue that it is another
lensed image of the same background QSO, and examine the intrinsic
properties of the lensed QSO based on the lens model.  We especially
try to constrain the effects of differential magnification since they
could introduce a significant distortion in the observed spectral
energy distribution (SED) as shown by Blain (1999).

After the initial submission of this paper, we were informed by Drs.\
Ibata and Lewis that a similar work was going to be submitted based on
the analysis of HST/NICMOS images of APM 08279+5255 (Ibata 1999).  The
two sets of observational data are in excellent agreement.  Both
studies (1) detect a third image, (2) derive similar positions and
relative brightnesses for all three images, and (3) conclude that the
third image is likely to be another lensed image of the same
background QSO and construct a similar lens model based on this
assumption.  Although our work was carried out independently of
their work, we have opted to incorporate their results whenever they
provide critical pieces of information to understand this object.
\section{OBSERVATIONS AND DATA REDUCTION}

\subsection{Near-IR Imaging}
The $K$-band images of APM 08279+5255 were taken on the night of UT
1998 October 3 with the Near Infra-Red Camera (NIRC;
\cite{Matthews94}) on the Keck I telescope on Mauna Kea in Hawaii.
NIRC uses a Hughes-SBRC 256 $\times$ 256 InSb array as the detector,
and is attached to the f/25 forward Cassegrain focus of the telescope,
producing a pixel scale of 0\farcs15/pixel with a field of view of
38\arcsec\ on a side.  For this observation, the NIRC image converter
(\cite{Matthews96}) was used with NIRC to achieve a high spatial
resolution.  The image converter changes the beam from f/25 to f/180,
producing a pixel scale of 0\farcs0206/pixel with a field of view of
5\farcs3 on a side.  For $K$-band imaging, the integration time was 10
seconds per image.  The highest resolution image was produced by
combining the four images (40 seconds total) with the minimum image
size (i.e., selective shift \& add).  The extremely good seeing that
night resulted in a FWHM of 0\farcs15 in this $K$-band image.

\subsection{Mid-IR Imaging}
The 12.5 $\mu$m and 17.9 $\mu$m images were initially taken on the
night of UT 1998 October 3 using the MIRLIN mid-IR camera
(\cite{Ressler94}) on the Keck II telescope.  MIRLIN uses a Boeing 128
$\times$ 128 Si:As array, and is attached to the f/40 bent Cassegrain
visitor port of the telescope, producing a pixel scale of
0\farcs138/pixel with a field of view of 17\arcsec on a side.  The
total integration time was 15 minutes at each wavelength.  These
images were used to measure the mid-IR flux density of APM 08279+5255.
Calibration was done by observing bright stars tied to the IRAS
calibration at 12 and 25 $\mu$m (IRAS Explanatory Supplement 1988).

The higher spatial-resolution 12.5 $\mu$m image of APM 08279+5255 was
taken on the night of UT 1999 November 25 using the Long Wavelength
Spectrometer (LWS; \cite{Jones93}) on the Keck I telescope in the
imaging mode.  LWS uses a Boeing 128 $\times$ 128 Si:As array, and is
attached to the f/25 forward Cassegrain focus of the telescope,
producing a pixel scale of 0\farcs08/pixel with a field of view of
10\farcs2 on a side.  The seeing was 0\farcs4 FWHM, and the total
integration time was 27 minutes.

\subsection{Near-IR Photometry}
Near-IR photometry was obtained on UT 1998 November 3 with the InSb
camera on the 200-inch Telescope at Palomar Observatory.  Calibration
was done by measuring stars listed by Elias et al. (1982).

The magnitudes and flux densities reported here are listed in
Table~\ref{new_flux}.  All the photometry was done with a
4\arcsec-diameter beam.
\section{RESULTS}

\subsection{The \mbox{\boldmath $K$}-band Image}
\begin{figure}
\centerline{\psfig{file=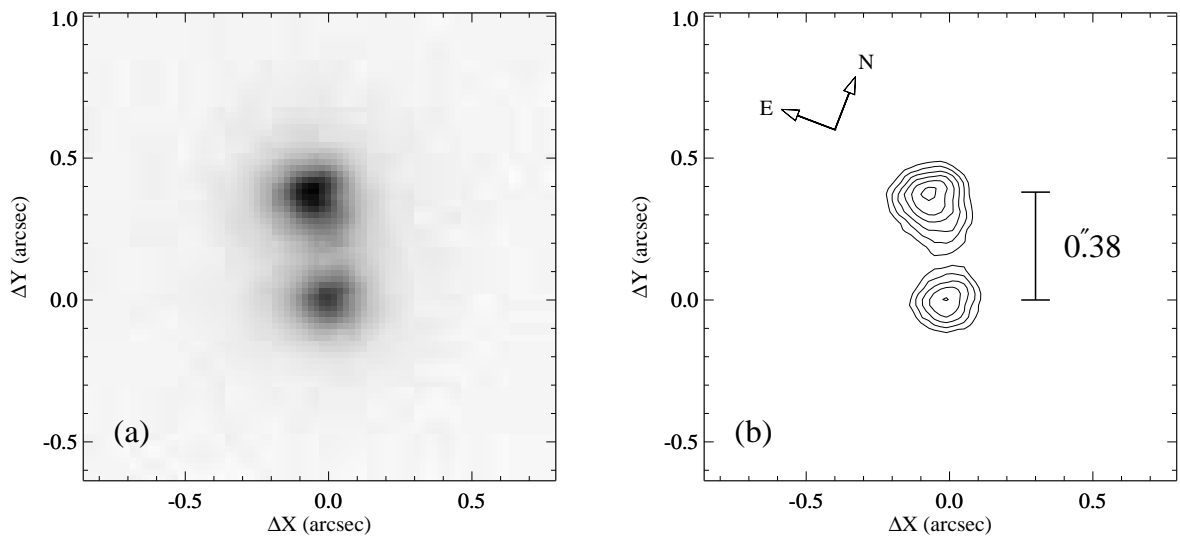,angle=90,width=6.5in}}
\figcaption[fig1.ps]{
(a) The $K$-band image of APM 08279+5255 taken with NIRC with a pixel
scale of 0\farcs02/pixel.  The seeing was 0\farcs15 FWHM.  The
coordinate origin was taken at the position of the lower image.  The
north is 21\arcdeg\ clockwise from the vertically up direction (see
the arrows in panel (b)). We chose not to rotate the image to
avoid any smoothing.  This orientation is adopted for all the
subsequent images/contours in this paper except for
Figure~\ref{mirlin_image}, whose orientation differs by
3\arcdeg ; (b) A contour map of (a).  The lowest contour is 18
$\sigma$ above the sky, and the subsequent contour levels increase by
a factor of 1.25.
\label{nirc}} 
\end{figure}

The $K$-band grey-scale image with scale 0\farcs0206/pixel is shown in
Figure~\ref{nirc}a with the contour map of the same data in
Figure~\ref{nirc}b.  It is immediately clear from this image that the
$K$-band image consists of two bright components, which is consistent
with the previous observations by Irwin et al. (1998) and Ledoux et
al. (1998).  The northern brighter source is slightly extended to the
south-west.  Since the southern source is quite circularly symmetric, 
this suggests that the extension in the northern source 
is a third fainter component.

We performed a point-spread-function (PSF) fitting using DAOPHOT
(\cite{Stetson87}) on the $K$-band image.  First, it was assumed that
the two main bright components are basically point sources, and that
we could construct the PSF by using the upper half of the upper
component and the lower half of the lower component.  Two axisymmetric
PSFs were first produced from two half PSFs by self-reflection, and
their average was taken to produce the final PSF.  In other words, we
assumed that the PSF is symmetric with respect to the horizontal axis
in Figure~\ref{nirc}.

To determine the position of the third component accurately enough as
the starting point for the PSF fitting, we first performed the two-PSF
subtraction.  The position of the third component was determined in
the residual image, and with this added information, we then performed
the three-PSF subtraction.

\begin{figure}
\vspace*{-3cm}
\psfig{file=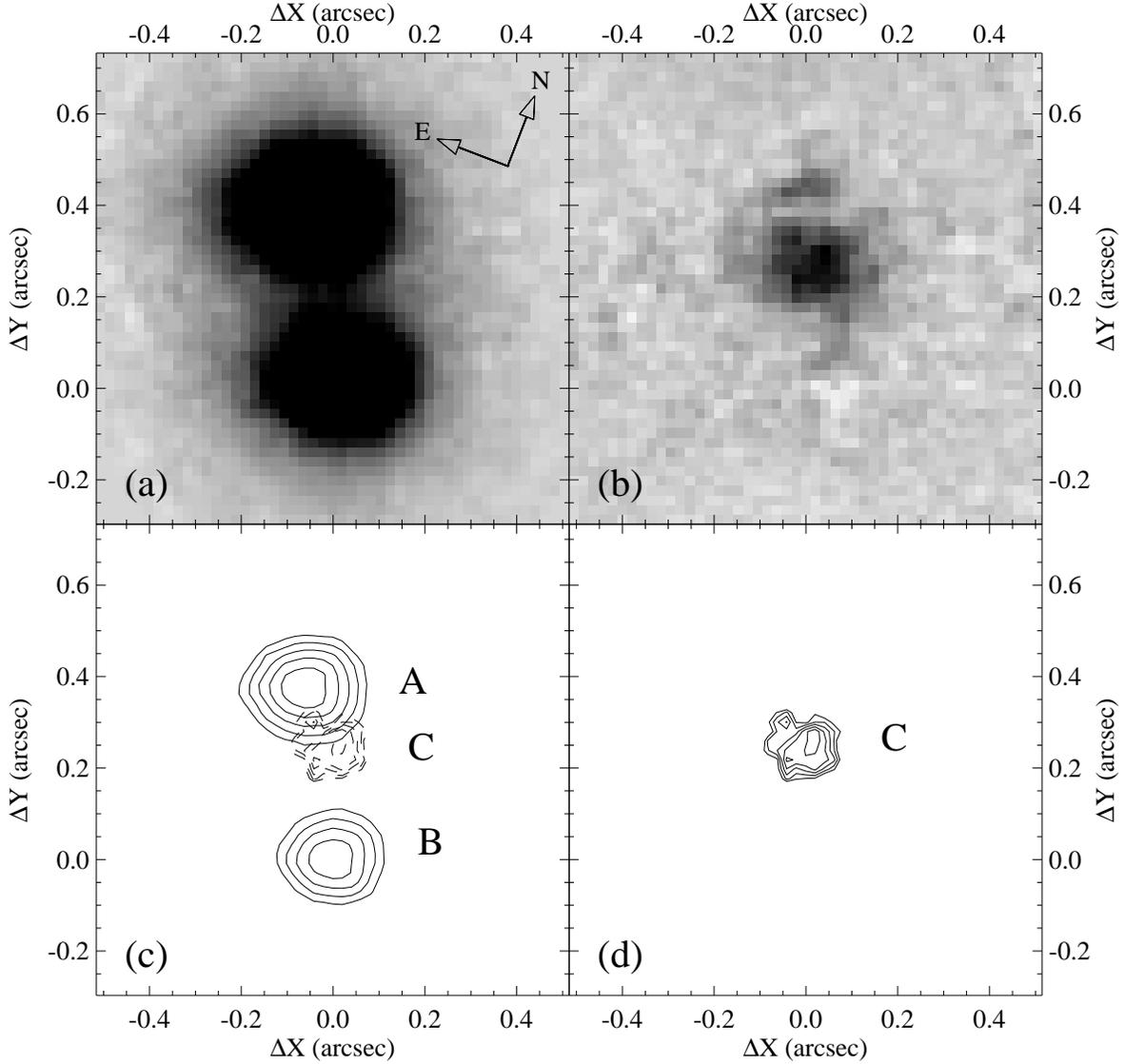,width=6.5in}
\vspace*{-3cm}
\caption[fig2.ps]{
(a) The $K$-band image of APM 08279+5255 showing components A and B.
Component C was subtracted from Figure~\ref{nirc}a.  The grayscale is
the same as that of (b) so that the relative brightness can be
compared directly.  The coordinate origin was taken at the position of
component B; (b) The $K$-band image showing component C.  Components A
and B shown in (a) were subtracted from Figure\ref{nirc}a; (c) A
contour map of (a).  The lowest contour level is 18$\sigma$ above the
sky, and the subsequent contour levels increase by a factor of 1.25.
Component C, shown in (d), is overlaid; (d) A contour map of (b).  The
lowest contour level is 9 $\sigma$ above the sky, and the subsequent
contour levels increase by a factor of 1.1.
\label{nirc_image}}
\end{figure}

The three components we have finally found are shown in
Figure~\ref{nirc_image}a and \ref{nirc_image}b together with their
contour maps Figure~\ref{nirc_image}c and \ref{nirc_image}d.  We refer
to these components as A, B, and C in descending order of
brightness.  The positions and relative brightnesses determined from
the PSF fitting for each component are listed in Table~\ref{imconfig}.
Using the flux ratios, the $K$ magnitudes of individual components
were derived from magnitude of $K=12.08$ mag for the total system.  The
derived FWHMs indicate that the seeing was 0\farcs15 when the image
was taken.  The FWHM of the component C indicates that this component
is also point-like.  For our further discussions, we assume that the
$K$-band image of APM 08279+5255 consists of three point sources.

Overall our measured positions and relative brightnesses of the three
components are in good agreement with the values derived by Ibata et
al. (1999). 

\subsection{The 12.5 \mbox{\boldmath $\mu$}m Image}
\begin{figure}
\vspace*{-3cm}
\centerline{\psfig{file=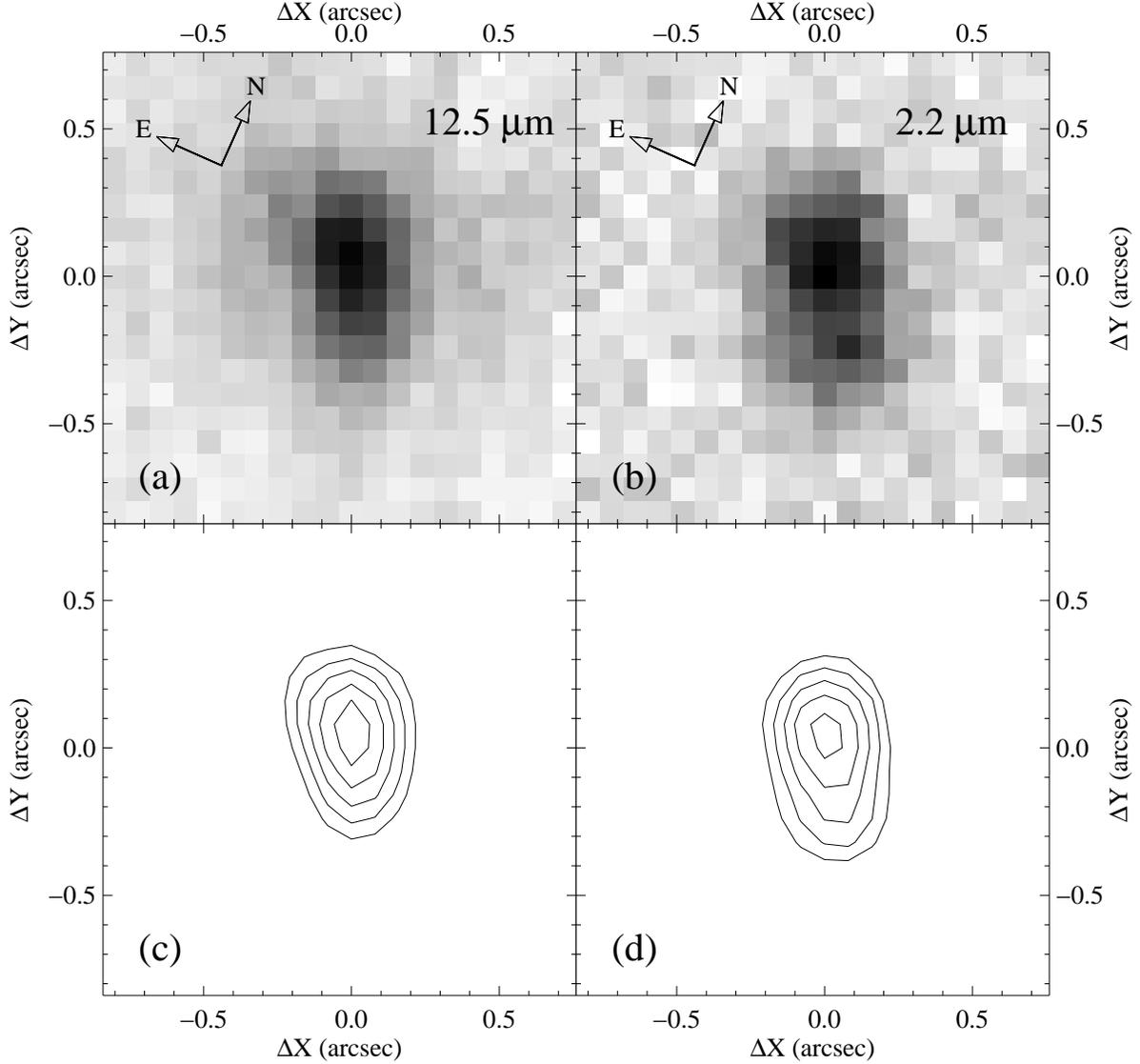,width=6.5in}}
\vspace*{-3cm}
\figcaption[fig3.ps]{(a) The 12.5 $\mu$m image of APM 08279+5255 taken
with LWS.  The pixel scale is 0\farcs08/pixel.  The seeing is
estimated to be FWHM $\sim$ 0\farcs4.  The coordinate origin was taken
arbitrarily.  North is 24\arcdeg\ clockwise from the vertically up
direction, 3\arcdeg\ different from that in the $K$ band image; (b) An
artificial image constructed from the three components detected in the
$K$-band.  Three point sources with the relative positions and fluxes
listed in Table~\ref{imconfig} were convolved with a Gaussian with a
FWHM$=$0\farcs4, and were sampled with a pixel scale of
0\farcs08/pixel.  Then, a Gaussian noise with the level seen in the
LWS image (a) was added.  The orientation is same as that of (a), and
the coordinate origin was taken arbitrarily; (c) \& (d) Contour maps
of (a) and (b) respectively.  The contour levels correspond to 40, 50,
60, 70, 80, and 90 \% of the peak level in each image (the sky level
is subtracted).
\label{mirlin_image}}
\end{figure}

The LWS 12.5 $\mu$m image of APM 08279+5255 is shown in
Figure~\ref{mirlin_image}a.  Figure~\ref{mirlin_image}b is an
artificial image showing how the three components detected in the
$K$ band would look like if observed with the LWS pixel scale
(0\farcs08/pixel) under the same seeing condition (FWHM $=$ 0\farcs4).
This image was simulated using the parameters listed in
Table~\ref{imconfig}.  Contour maps of the images are also shown in
Figure~\ref{mirlin_image}c and \ref{mirlin_image}d, respectively.

The morphological resemblance is clear.  Although there might be a
small difference in morphology between the two images, we cannot say
with confidence that it is real.  The 12.5 $\mu$m image shape changes
from image to image considerably because of the lower signal-to-noise
ratio, and the image shape might have been smeared when a large number
of images were combined.  Therefore, based on the overall ellipticity
of the image contours, we conclude that the image configuration at
12.5 $\mu$m is similar to that in the $K$ band.
\section{DISCUSSION}
 
\subsection{The Lens Model}

To construct a gravitational lens model, we use the elliptical
effective lensing potential $\psi$ in the following form (Blandford
\& Kochanek 1987; Narayan \& Bartelmann 1999):
\begin{equation}
  \psi(\theta_{1},\theta_{2}) = \theta_{E} [\theta_{c}^{2} +
    (1-\varepsilon)\theta_{1}^{2} +
    (1+\varepsilon)\theta_{2}^{2}]^{1/2}, \label{epot}
\end{equation}
where $\theta_{E}$ is the Einstein radius, $\theta_{c}$ and
$\varepsilon$ are the core radius and ellipticity of the lensing
potential, and $\theta_{1}$ and $\theta_{2}$ are rectangular
coordinates in radians with respect to an arbitrarily defined optic
axis.  Although we treat the Einstein radius itself as a free
parameter, it can also be expressed as,
\begin{equation}
  \theta_{E} = \frac{D_{ls}}{D_{s}} 4\pi \frac{\sigma^{2}_{v}}{c^2},
  \label{re}
\end{equation}
where $D_{ls}$ is the standard angular distance between the lens and
the source (cf., \cite{Fukugita92}), $D_{s}$ is the angular distance
from the observer to the background source, $\sigma_{v}$ is the
internal velocity dispersion of the lensing galaxy (i.e., the mass),
and $c$ is the speed of light.  For a given Einstein radius, this
equation sets the constraints on the mass and redshift of the lens.
In addition to these parameters specifying the shape of the potential,
the center position ($x_{l}$,$y_{l}$) and the position angle
($\gamma$) of the lensing potential must be specified.  All together, the
model contains six unknown parameters ($\theta_{E}$, $\theta_{c}$,
$\varepsilon$, $\gamma$, $x_{l}$ and $y_{l}$).

The major uncertainty with APM 08279+5255 is the nature of component
C.  There are two possibilities: it is either a third lensed image of
the same background source or the lensing galaxy.  We will construct
models based on both these assumptions, and evaluate their validity
based on the available observational data.

\subsubsection{Three-Image Model}
\begin{figure}
\vspace*{-3cm}
\centerline{\psfig{file=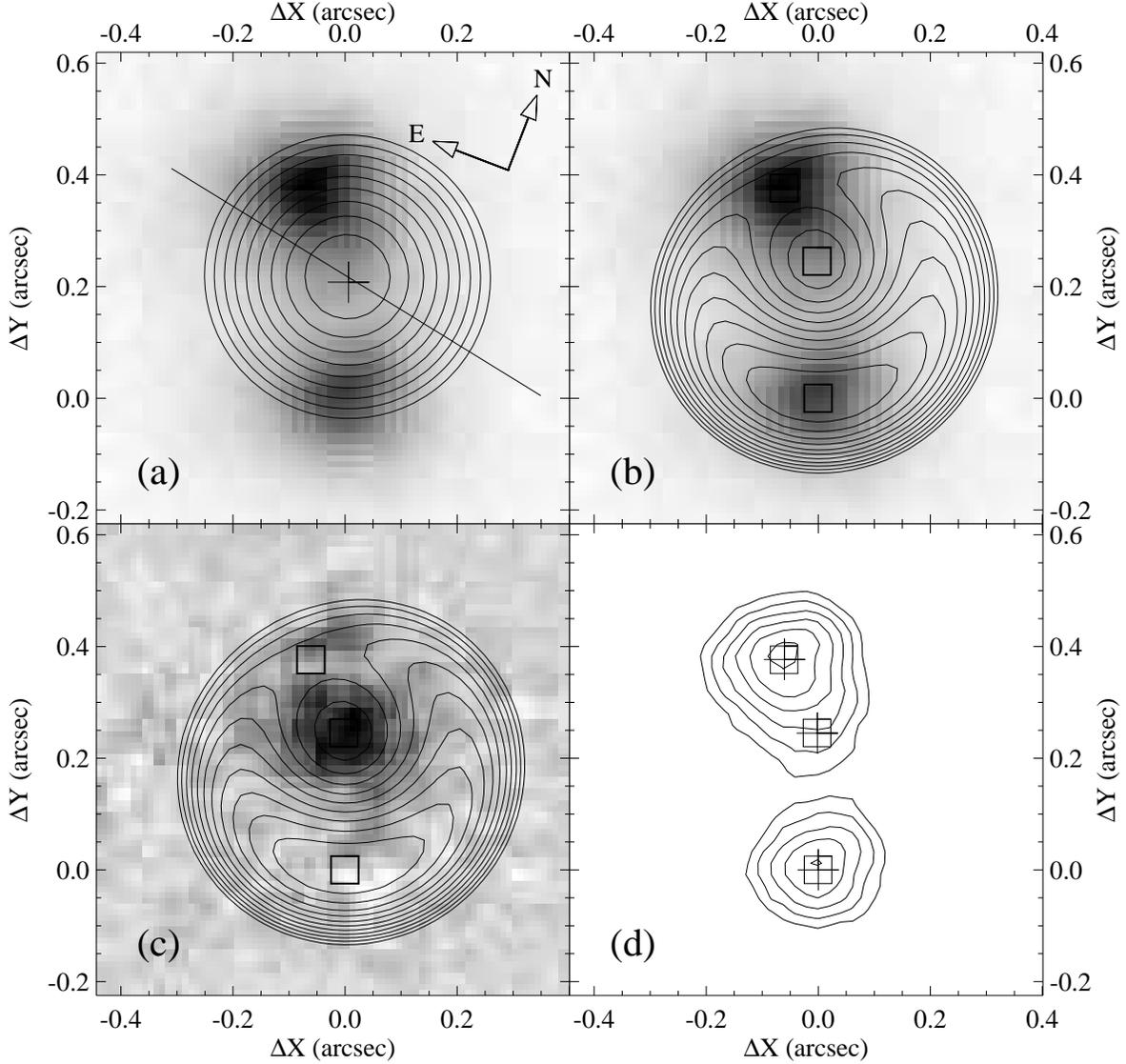,width=6.5in}}
\vspace*{-3cm}
\figcaption[fig4.ps]{(a) A contour plot showing the effective lensing
potential of the three-image model overlaid on the $K$-band image
(Figure~\ref{nirc}a).  The cross denotes the position of the
background source while the solid line indicates the direction of the
major axis of the potential.  The coordinate origin was taken at the
position of the component B; (b) The time-delay surface overlaid on
the $K$-band image.  The three squares indicate the expected positions
of the lensed images based on the model; (c) The same time-delay surface
overlaid on Figure~\ref{nirc_image}b, which contains the component C
only; (d) The contour map from Figure~\ref{nirc}b with the three image
positions measured in the $K$-band image (crosses) and those derived
from the model (squares).
\label{lens_model}}
\end{figure}

Three lensed images provide six constraints: two relative brightnesses
and four coordinates giving two relative image positions with respect
to the other.  Since the number of model parameters is also six,
we can determine the values of the parameters, but cannot assess the
goodness of the fit.  The best-fit model was searched by taking
component B as the reference and varying the parameters such that the
source positions of components A and C coincide with that of B while
the derived relative magnifications approach the observed flux ratios.
Our best-fit model parameters are shown in Table~\ref{bestfit}.
Figure~\ref{lens_model} shows the profile of the effective lensing
potential, the time-delay surface, and the expected image positions.
Basically, this model reproduces the positions and relative
brightnesses of three images well within the observational
uncertainties.  The lens is almost round ($\varepsilon \sim 0.01$) and
has a large core radius ($\theta_{c} \sim 0\farcs2$).  This is because
the three images are almost in a straight line (i.e., small
$\varepsilon$) and the third image is very bright (i.e., large
$\theta_{c}$).  The total magnification factor for a point
source\footnote{We originally derived a value of 71, which was quoted
by Ibata et al. (1999), but our later calculation increased the value
to 86, which agrees with Ibata et al.'s value of 90.}  was
calculated to be 86.

\subsubsection{Two-Image Model}
Ibata et al. (1999) presented the possibility that component C might
be the lensing galaxy rather than a third lensed image.  In this case,
the number of the model parameters reduces to three ($\theta_{e}$,
$\varepsilon$, and $\gamma$).  The core radius $\theta_{c}$ can be set
to 0 because the lack of a third image implies a singular potential
core while the position of component C directly determines the
position of the lensing galaxy ($x_{l}$, $y_{l}$)\footnote{Our model
is slightly different from that of Ibata et al. (1999) in that we fix
the lens position at the position of component C.  Their model treats
the lens position as a free parameter.}.  On the other hand, two
lensed images provide three constraints: one relative brightness and
two coordinates giving one relative image position with respect to the
other.  Again, the number of the model parameters and that of the
constraints are the same.  The derived parameters are shown in
Table~\ref{bestfit}.  The major difference from the three-image model
is the much lower value of total magnification (7).  Also, the
ellipticity of the potential is becoming significant (0.08).  The
overall structure of the time-delay surface is very similar to that of
the three-image model.

\subsection{Magnification of Extended Sources}
The background source is likely to be spatially extended at longer
wavelengths.  Therefore, it is necessary to understand how the lens
distorts and magnifies an extended source as the outer edge of the
source approaches/crosses the caustics.  For a background source that
is a uniform circular disk, the lensed image can be constructed with
the methods of Schramm \& Kayser (1987).  Since the surface brightness
is preserved in gravitational lensing, the areal ratio between the
source and the images gives the magnification factor.

\begin{figure}
\centerline{\psfig{file=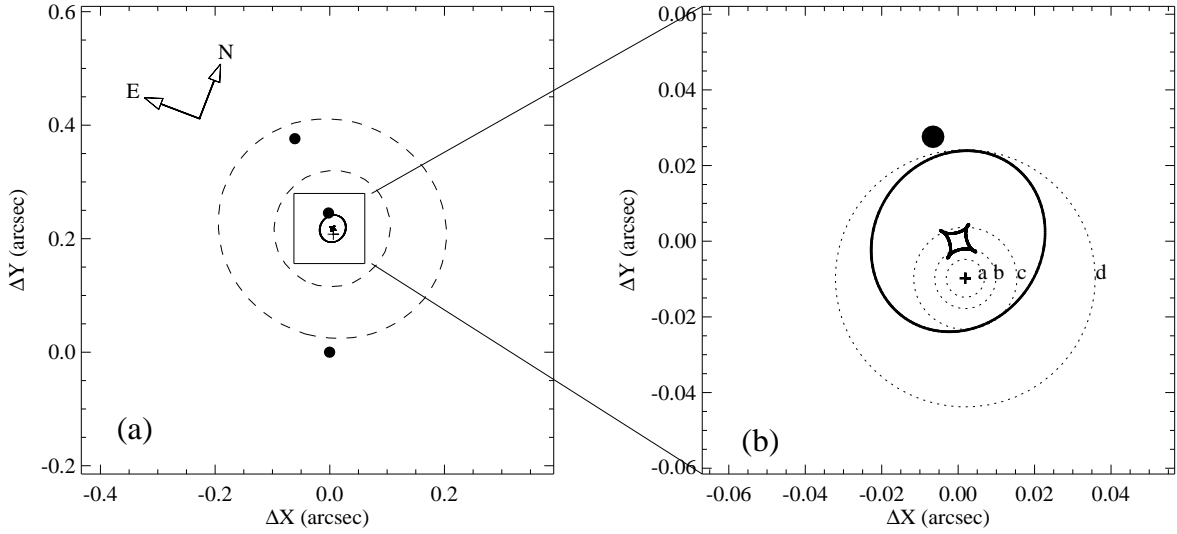,angle=90,width=6.5in}}
\figcaption[fig5.ps]{Three-image model: 
(a) A plot showing two critical curves (two outer dashed
lines), two caustics (two inner solid lines in the square), the
background QSO position (cross), and three lensed images (solid
circles).  The coordinate origin was taken at the position of the
component B; (b) A plot magnifying the square region in (a).  
The dotted lines show the shapes of four extended sources (i.e., constant surface
brightness disks) with different radii.  The source radii of a, b, c,
and d correspond to those of the panels (a), (b), (c), and (d) in
Figure~\ref{model_extend} and \ref{mag_extend}, respectively.  The
coordinate origin was taken at the center of the lensing potential.
\label{caustic}}
\end{figure}

\begin{figure}
\centerline{\psfig{file=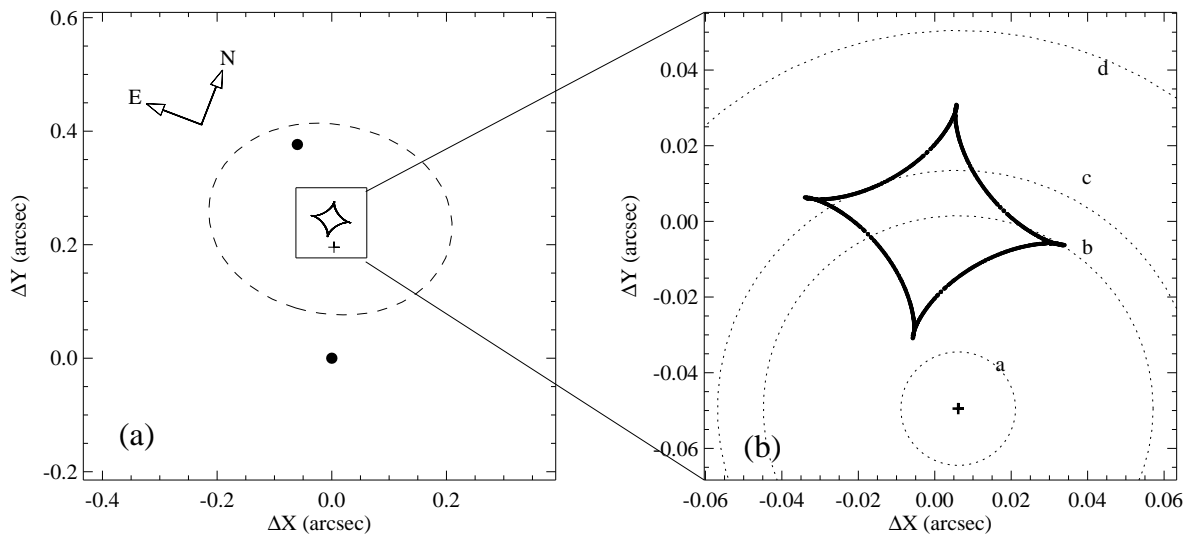,angle=90,width=6.5in}}
\figcaption[fig6.ps]{Two-image model:
(a) This plot is same as Figure~\ref{caustic} except that it is for
the two-image model.  Because of the singularity at the core (i.e.,
$\theta_{c}=0$), there is only one caustic and one corresponding
critical curve; (b) A plot magnifying the square region in (a).  The
dotted lines show the shapes of four extended sources (i.e., constant
surface brightness disks) with different radii.  The source radii of
a, b, c, and d correspond to those of the panels (a), (b), (c), and
(d) in Figure~\ref{model_extend2} and \ref{mag_extend}, respectively.
The coordinate origin was taken at the center of the lensing potential.
\label{caustic2}}
\end{figure}

\begin{figure}
\vspace*{-3cm}
\centerline{\psfig{file=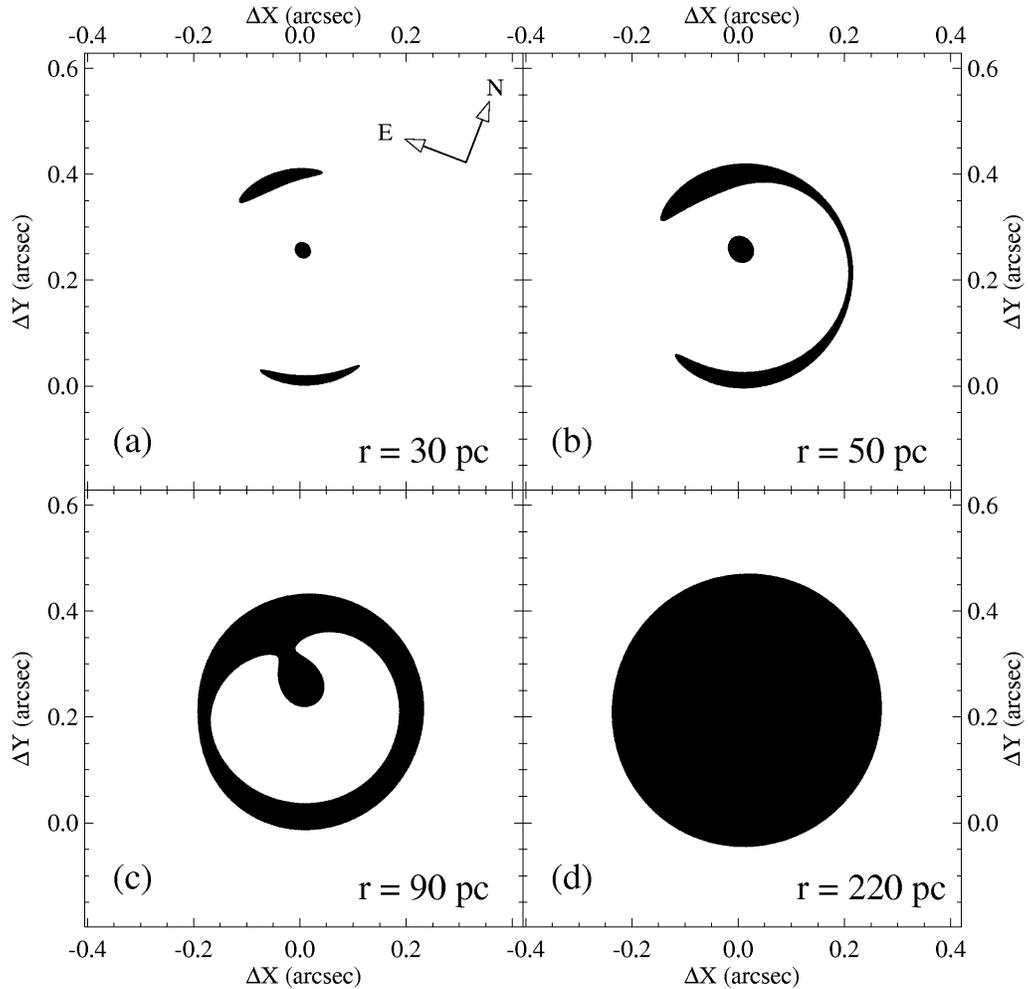,width=6.5in}}
\vspace*{-3cm}
\figcaption[fig7.ps]{The gravitationally lensed images of an extended source
produced by the three-image model.  The intrinsic source was assumed
to be a disk with a constant surface brightness.  The radius of the
source was assumed to be (a) 0\farcs005 (30 pc), (b) 0\farcs008 (50
pc), (c) 0\farcs0135 (90 pc), and (d) 0\farcs034 (220 pc),
respectively.  The surface brightness of the image is preserved by the
gravitational lensing process.  The coordinate origin was taken at the
position of the component B.
\label{model_extend}}
\end{figure}

\begin{figure}
\vspace*{-3cm}
\centerline{\psfig{file=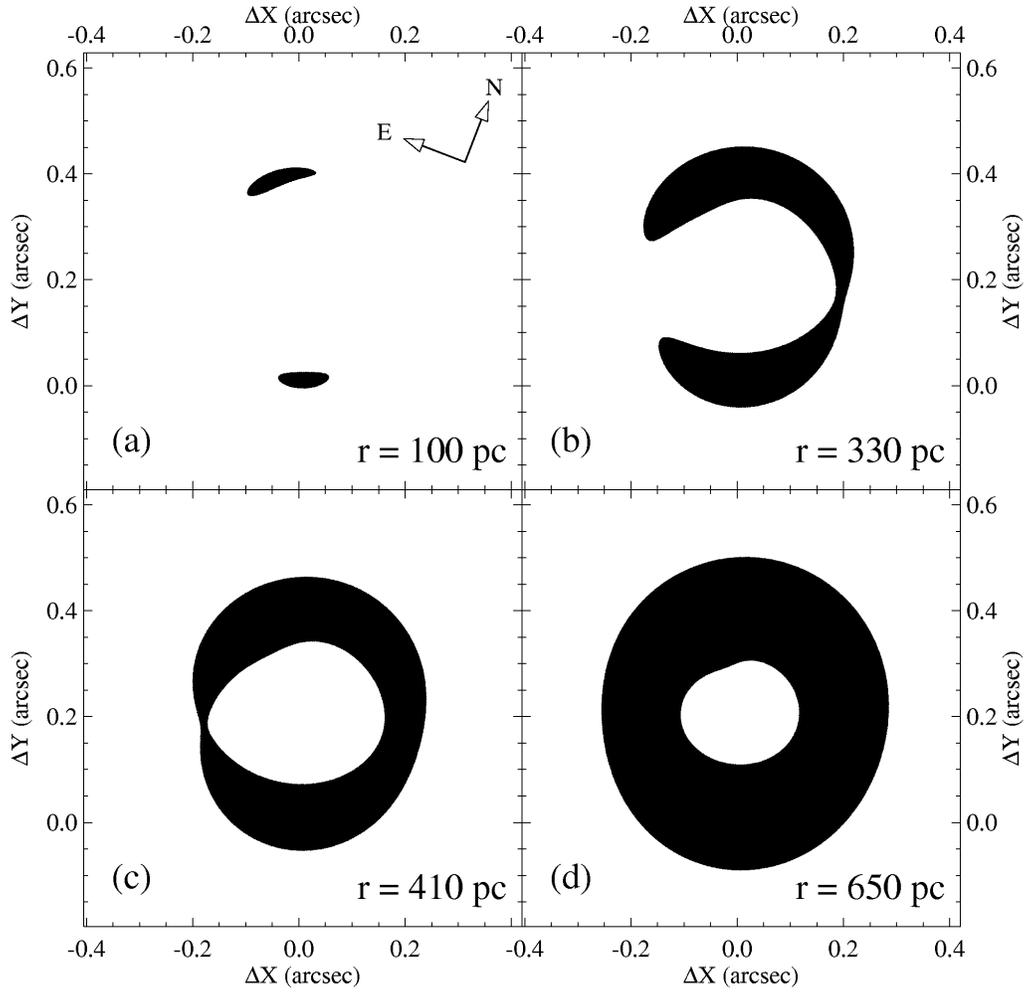,width=6.5in}}
\vspace*{-3cm}
\figcaption[fig8.ps]{This plot is same as Figure~\ref{model_extend}
except that it is for the two-image model.  The radius of the source
was assumed to be (a) 0\farcs015 (100 pc), (b) 0\farcs051 (330 pc),
(c) 0\farcs063 (410 pc), and (d) 0\farcs1 (650 pc), respectively.
\label{model_extend2}}
\end{figure}

\begin{figure}
\centerline{\psfig{file=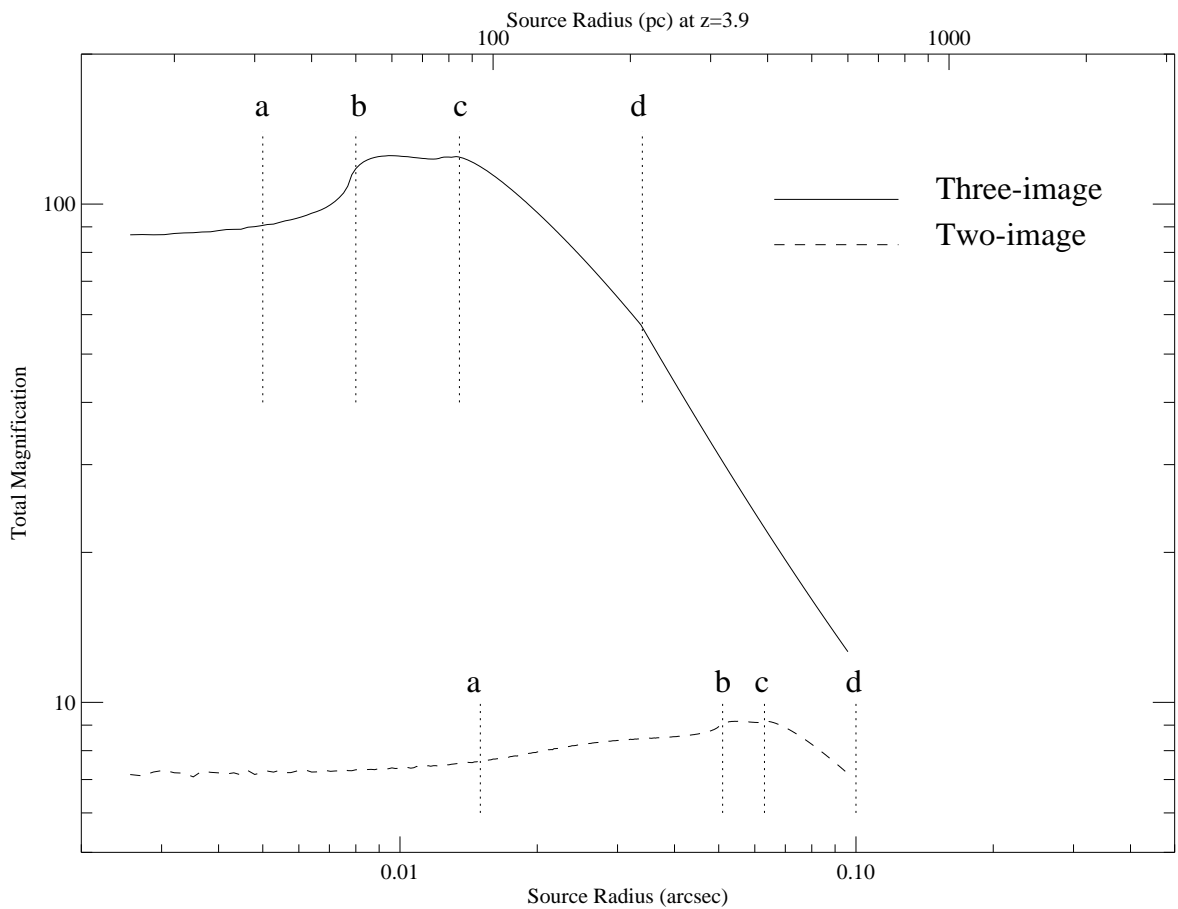,angle=90,width=6.5in}}
\figcaption[fig9.ps]{The total magnification produced by the lens
models as a function of the source radius.  The source is assumed to
be a disk with a constant surface brightness.  The solid curve shows
the magnification factor of the three-image model while the dashed
curve shows that of the two-image model.  The vertical dotted lines
indicate the source radii of the panels (a), (b), (c), and (d) in
Figure~\ref{caustic} and \ref{model_extend} for the three image-model,
and Figure~\ref{caustic2} and \ref{model_extend2} for the two-image
model.
\label{mag_extend}}
\end{figure}

For the two models, we show how the source crosses the caustics
(Figures~\ref{caustic} and \ref{caustic2}) and how the shapes of the
lensed images change (Figure~\ref{model_extend} and
\ref{model_extend2}).  The behavior is similar in the two models: as
the source size increases, it first forms an arc (b) and later turns
into a ring (c).  In the case of the three-image model, the ring
quickly becomes a filled disk due to the third image while the ring is
not completely filled in the two-image model even with a source radius
of 650 pc (d).  Figure~\ref{mag_extend} shows how the total
magnification factor changes as the source radius increases in the two
models.

There are three major differences between the two models in terms of
their response to extended sources.  First, a much larger source size
is required in the two-image model for each transition of the lensed
image shape as shown in Figure~\ref{model_extend2}.  Second, the
magnification is much greater in the three-image model.
Figure~\ref{mag_extend} shows that the magnification factor of the
three-image model could be as large as $\sim 120$ when the source
radius is 50--90 pc while that of the two-image model is more than an
order of magnitude less ($\sim 7$) for the same radius range.  Third,
the magnification factor of the two-image model hardly changes with an
increasing source size while the magnification factor of the
three-images model is very sensitive to such a change
(Figure~\ref{mag_extend}).  In other words, the effects of
differential magnification could be significant with the three-image
model.

\subsection{Differential Reddening}
One uncertainty underlying the discussion so far is differential
reddening.  For example, if the three-image model is correct, the line
of sight to component C intersects the lensing galaxy at a point only
$\sim$ 0\farcs03 from the galaxy's center (Figure~\ref{caustic}b).
Such an angular separation corresponds to 200--250 pc at a lensing
galaxy's redshift of $0.5<z<4$.  If this is the case, the brightness
of component C might be more heavily affected by the reddening in the
lensing galaxy than those of the other two components.  This would
produce observed flux ratios considerably different from the intrinsic
ones.

However, the currently available data seem to indicate that
differential reddening is not significant in this system.  Ibata et
al. (1999) noted that the three components have almost identical
colors from 1.10 $\mu$m to 2.05 $\mu$m.  Our $K$-band image also do
not show any sign of change in the flux ratio or image morphology,
which would have resulted if component C brightens significantly at
longer wavelengths.  The 12.5 $\mu$m image also seems to be consistent
with the $K$-band image.  Therefore, it seems unlikely that the
derived models are in serious error due to differential reddening.

\subsection{The Nature of the Third Image}
As seen in Figure~\ref{mag_extend}, depending on whether component C
is a third lensed image or a lensing galaxy, the magnification factor
could be drastically different.  Therefore, the nature of component C
is a decisive factor when determining the intrinsic properties of APM
08279+5255.  We list three arguments favoring the idea that
component C is a third lensed image rather than the lensing galaxy.

\subsubsection{The large apparent brightness of component C}
If component C is the lensing galaxy, then it cannot be an ordinary
field galaxy.  The small separation between the lensed images
(0\farcs4) requires the lensing galaxy to be either a low-mass
low-redshift galaxy or a massive high-redshift galaxy in order not to
split the lensed images too far apart.  However, neither type of
galaxy with a normal mass-to-light ratio could produce such a bright
apparent magnitude as $K=14.5$ mag.  More quantitatively, inserting
$\theta_{E}=0$\farcs2 in equation~(\ref{re}), we obtain possible
combinations of the lensing galaxy redshift and velocity dispersion
$(z_{l}, \sigma_{v} ({\rm km/s}))$ as (0.5, 100), (1, 120), (2, 170),
and (3, 280).  To illustrate how much the lensing galaxy must be
overluminous, if we take an L$^{*}$ galaxy (i.e., $\sigma \sim 200$
km/s) as the lens, it implies that $z_{l} \sim 2$, but the expected
observed $K$ magnitude of such a galaxy would be $K \sim 20$ mag.
Therefore, it requires such a lensing galaxy to be overluminous by
more than five magnitudes.  Together with the evidence that component
C is point-like, this would require that component C is a luminous
AGN/QSO nucleus of the lensing galaxy.  Although such a QSO--AGN/QSO
gravitational lensing may happen (Gott \& Gunn 1974), the probability
for such an event is in general very small.

\subsubsection{Dark Ly$\alpha$-forest line cores}
The high-resolution Keck HIRES spectrum of APM 08279+5255
(\cite{Ellison99a,Ellison99b}) shows no residual flux in the core of
strong Ly$\alpha$ forest lines up to an observed wavelength of 5715
\AA.  If component C is a lower-redshift AGN/QSO, its continuum flux
should easily be detectable in the core of these strong Ly$\alpha$
lines, given the high signal-to-noise ratio of this HIRES spectrum.
This means that if component C is an AGN/QSO, its redshift must be
larger than 3.7, and it seems rather contrived if a $z=3.9$ QSO is
gravitationally lensed by another QSO system at $z>3.7$.

\subsubsection{Color}
As already mentioned in the discussion of differential reddening,
components A, B, and C have almost identical colors from 1.1 $\mu$m to
12.5 $\mu$m.  This is easy to understand if component C is also a
lensed image of the same background source.

Based on these arguments, we conclude that component C is a third
lensed image of the QSO.  Therefore, we investigate the intrinsic
properties of APM 08279+5255 based on the three-image model, although
in the end this question can be settled with spectroscopy of
component C.  If, however, the two-image model is correct, its
consequences are easy to derive because this model is not sensitive to
differential magnification: the magnification factor is $\sim$ 7-10
over the relevant spectral range, and therefore APM 08279+5255 must be
an extremely luminous object with the intrinsic bolometric luminosity
$\sim 5 \times 10^{14}$ L$_{\odot}$.  Also, the shape of the
intrinsic SED must be close to what is observed.

\subsection{Differential Magnification}

In this section, we discuss the consequences for differential
magnification of our preferred three-image model.  Unlike the
two-image model, the magnification factor of the three-image model is
sensitive to the source size (Figure~\ref{mag_extend}).  Therefore,
the effects of differential magnification must be evaluated before the
intrinsic properties of the lensed QSO can be discussed.  For the
discussions below, references are made with respect to the restframe
wavelength of APM 08279+5255.

\subsubsection{Magnification in the Restframe UV/Optical}
The fact that three lensed images are completely point-like in the
restframe $B$ band (observer's $K$ band) sets the upper limit of
$\sim$ 20 pc on the source radius.  If the source radius were larger
than 20 pc, we should see the effects of image elongation seen in
Figure~\ref{model_extend}a.  In the restframe UV, the source size is
expected to be either comparable to or smaller than that in the
restframe optical.  From Figure~\ref{mag_extend}, it can be seen that
the magnification factor corresponding to this range of source size is
$\sim$ 90, which is same as the value derived by Ibata et al. (1999).

\subsubsection{Magnification in the Restframe Near-IR}
\begin{figure}
\vspace*{-3cm}
\centerline{\psfig{file=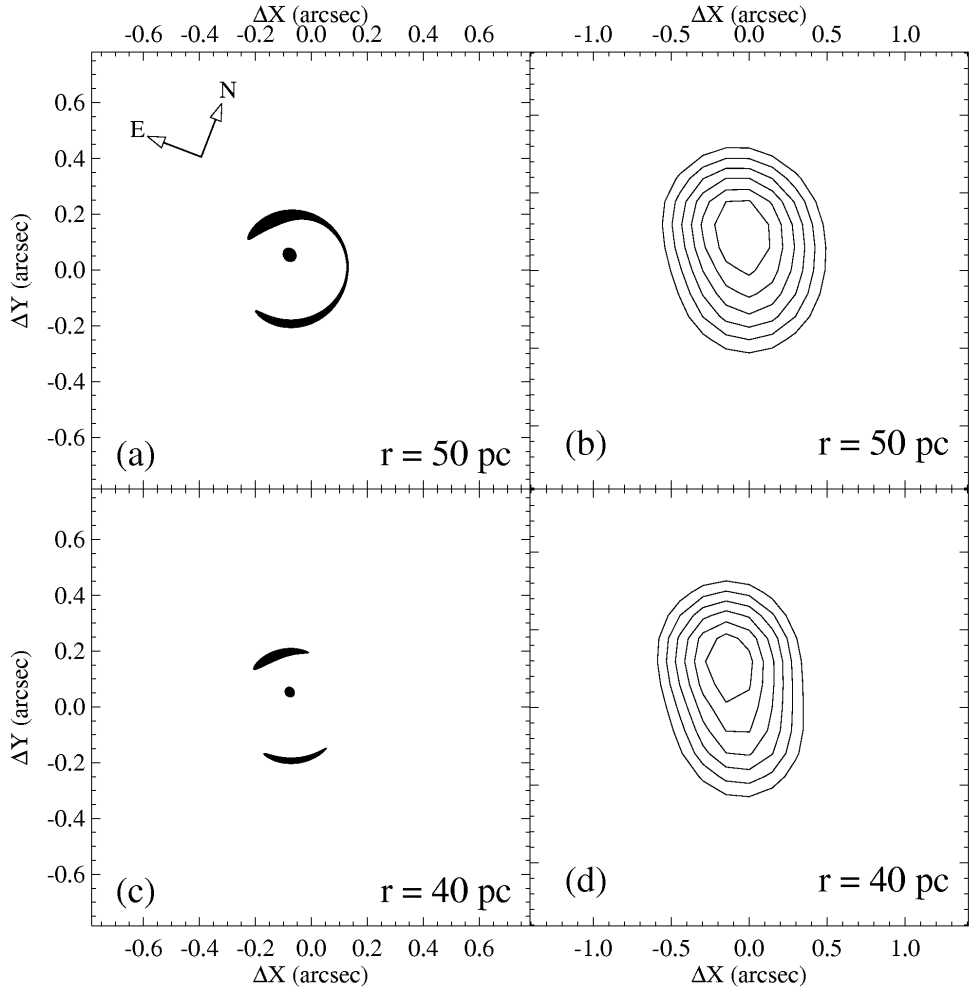,width=6.5in}}
\vspace*{-3cm}
\figcaption[fig10.ps]{(a) The expected lensed image of APM 08279+5255 
with an intrinsic source radius of 0\farcs008 (50 pc) in the
three-image model; (b) A contour map of (a) convolved with a Gaussian
PSF with a FWHM of 0\farcs4 and rebinned to a pixel scale of
0\farcs08/pix , simulating the LWS 12.5 $\mu$m image.  The contour
levels are 40, 50, 60, 70, 80, and 90 \% of the peak level (the sky
level is subtracted); (c) The expected lensed image with an intrinsic
source radius of 0\farcs006 (40 pc) in the three-image model; (d) A
contour map of the the image (c) convolved with a gaussian PSF with a
FWHM of 0\farcs4.  The contour levels are 40, 50, 60, 70, 80, and 90
\% of the peak level (the sky level is subtracted).  The difference of
ellipticity between (b) and (d) is clearly noticeable.
\label{model_12}}
\end{figure}

Since the spatial resolution of the restframe near-IR (observed 12.5
$\mu$m) image is not high enough to detect the morphology of
each component, it is not possible to set as stringent a limit on the
source radius as in the restframe optical.  As seen in
Figure~\ref{model_extend}, once the source radius reaches 50 pc,
components A and B will connect with each other and form an arc, and
this would produce a lower ellipticity in the resultant image
(Figure~\ref{model_12}).  However, this has not been seen in
Figure~\ref{mirlin_image}a and c.  Based on the overall ellipticity of
the restframe near-IR image, we conclude that the source radius in the
restframe near-IR (observer's 12.5 $\mu$m) must be less than 50 pc,
which corresponds to a magnification factor of 90--120
(Figure~\ref{mag_extend}).

\subsubsection{Magnification in the Restframe Mid-IR}
There exists no high-resolution spatial information in the restframe
mid-IR (observer's far-IR) which can be used to determine the
magnification factor directly.  However, it is possible to put
constraints on the magnification factor based on the restframe mid-IR
luminosity.

With the assumption that the mid-IR emitting region is intrinsically a
circular disk on the sky with a constant specific intensity, the
mid-IR SED can be modeled with the following expression:
\begin{equation}
  f_{\nu_{obs}} = m Q_{em}(\nu_{e}) B_{\nu_{e}}(T_{d})(1+z_{s})^{-3}
            \pi \left(\frac{r}{D_{s}}\right)^{2}.
     \label{obs_flux}
\end{equation}
Here, $m$ is the magnification factor, $Q_{em}$ is the dust
emissivity, $B_{\nu_{e}}$ is the Planck function evaluated at the
emitted frequency, $T_{d}$ is the dust temperature, $z_{s}$ is the
redshift of the background source, and $r$ is the intrinsic radius of
the emitting region.  The Planck function is evaluated at the emitted
frequency ($B_{\nu_{e}}$) while the observed flux density is expressed
at the observed frequency ($f_{\nu_{obs}}$;
$\nu_{e}=\nu_{obs}(1+z_{s})$).

As the dust temperature, we adopt the CO gas temperature of 200 K
derived by Downes et al. (1999) from the ratio of the CO (4--3) and
(9--8) lines.  If this 200 K region is strongly magnified, the dust
global temperature could be much lower than 200 K, but it does not
affect our discussion here because dust at a temperature significantly
below 200 K would not provide a significant flux in the restframe
mid-IR.

Only specific combinations of $m$ and $r$ are allowed by the lens
model (Figure~\ref{mag_extend}).  For each allowed combination, the
corresponding value of $Q_{em}$ is determined such that
equation~(\ref{obs_flux}) reproduces the observed IRAS 100 $\mu$m
flux.  Since 100 $\mu$m is close to the peak of a 200 K blackbody at a
redshift of 3.91, the value of $Q_{em}$ essentially scales the total
energy output of such a blackbody.  Therefore, its value can be
regarded as equivalent to that of the Planck-averaged dust emissivity
(Draine \& Lee 1984).

As long as the dust emissivity $Q_{em}$ is larger than $\sim$ 0.03, it
is always possible to find a set of a source radius and a
magnification factor which reproduces the observed 100 $\mu$m flux.
However, the emissivity cannot be smaller than this because the
intrinsic source size would become larger than 0\farcs3, which would
produce a gravitationally magnified image with a diameter $\sim$
1\arcsec, roughly the upper limit on the millimeter source size set by
Downes et al. (1999).  A Planck-averaged emissivity of 0.03 is similar
to that of astronomical silicate grains with a radius 0.01-1 $\mu$m at
200 K as found by Draine and Lee (1984).

Depending on the dust emissivity, the magnification factor in the
mid-IR could be anywhere between 4 ($Q_{em}=0.03$) and 120
($Q_{em}=1$), which corresponds to the source radius of 1.8 kpc and 60
pc, respectively.  The case of $Q_{em}=1$ corresponds to an optically
thick source of temperature 200 K.  The source radius of the 200 K
component becomes larger with a smaller dust emissivity because dust
grains can be heated to 200 K at a farther distance from the central
source.

\subsubsection{Magnification in the Restframe far-IR/submm}
The range of possible magnification factors in the restframe
far-IR/submm is the same as that in the mid-IR (i.e., 4--120).
However, it cannot exceed the magnification factor in the mid-IR since
the far-IR/submm emitting region is expected to be larger than the
mid-IR emitting region.  The magnification factor could be exactly the
same in the restframe mid-IR and far-IR/submm if the emission from the
200 K component also dominates in the latter wavelength regime.

Table~\ref{magtable} summarizes the derived magnification factors at
different wavebands and the corresponding source sizes.  It can be
seen that the effects of differential magnification are negligible
through the restframe UV, optical, and near-IR while it could
potentially be significant in the restframe mid-IR/far-IR/submm.

\subsection{Intrinsic Properties}

\subsubsection{Spectral Energy Distribution}
\begin{figure}
\centerline{\psfig{file=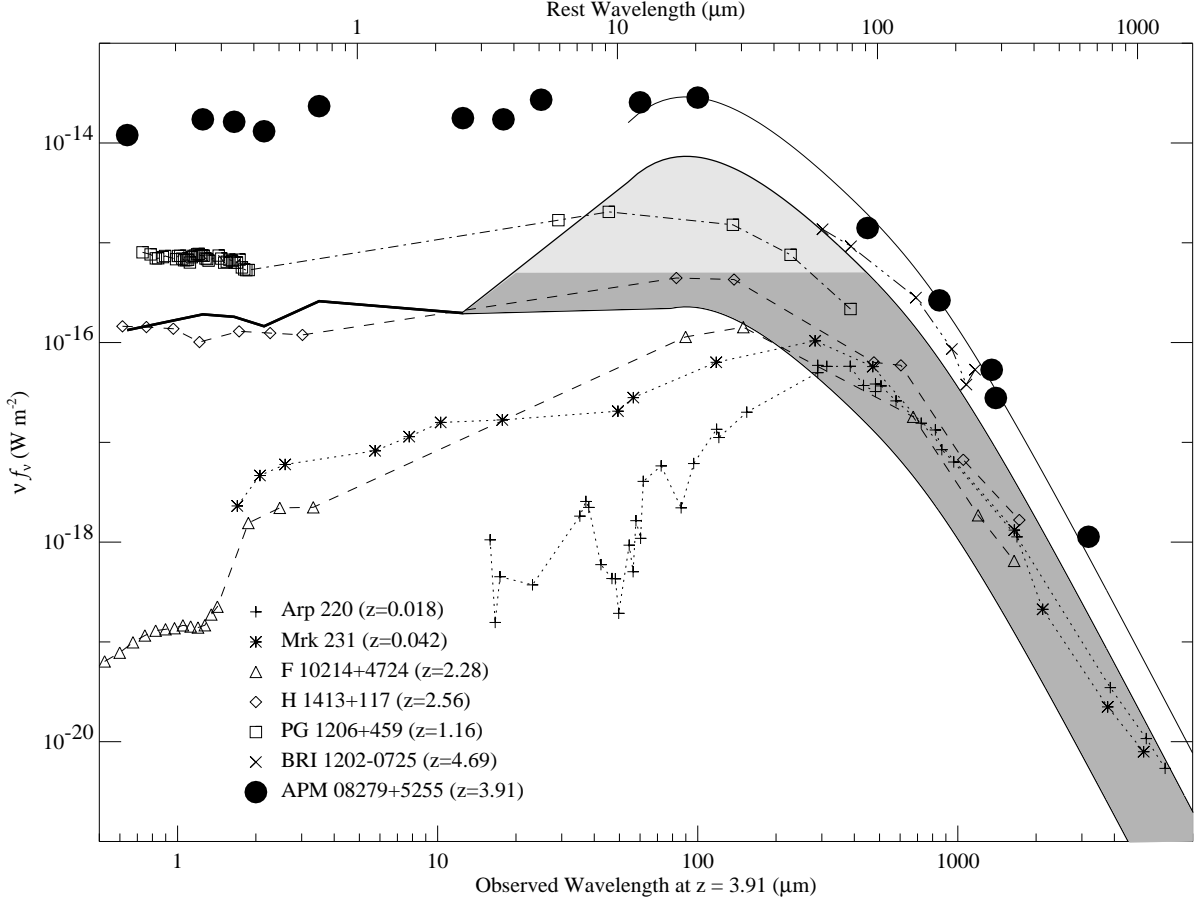,angle=90,width=6.5in}}
\figcaption[fig11.ps]{Comparison of the SEDs of various
ultra-/hyper-luminous galaxies/QSOs.  The SED of APM 08279+5255 was
downscaled by 90 at $\lambda_{obs} \leq 12.5 \mu$m (the thick solid
line).  The shaded area shows the allowed range for the intrinsic SED
of APM 08279+5255 at $\lambda_{obs}>12.5 \mu$m.  The area was
determined by demagnifying the empirical fit to the observed SED (the
solid line connecting the solid circles) by factors of 4 and 120.  The
data points at $\lambda_{obs} = 17.9$ and 25 $\mu$m are not used, and
the points at $\lambda_{obs} = 12.5$ and 60 $\mu$m are directly
connected.  The intrinsic SED is probably flat up to $\lambda_{rest}=
20 \mu$m (dark-shaded area; see text).  All the other SEDs are shifted
to a redshift of 3.91.  The SED of H1413+117 (Cloverleaf QSO) is
downscaled by a magnification factor of 10 (see Kneib et al. (1998)
for the most recent discussion of the magnification factor).  The SED
of F10214+4724 is downscaled by a factor of 100 at $\lambda_{rest} <
0.3 \mu$m (Eisenhardt et al. 1996), 50 at 0.3 $\mu$m
$<\lambda_{rest}<1 \mu$m (Evans et al. 1999), and 30 at
$\lambda_{rest} > 10 \mu$m (Eisenhardt et al. 1996).  The SED data
were taken from Barvainis et al. (1995), Benford et al. (1999),
Downes et al. (1999), Haas et al. (1998), Ibata et al. (1999), Irwin
et al. (1998), Isaak et al. (1994), Klaas et al. (1997), Lewis et
al. (1998), Neugebauer et al. (1987), Rigopoulou et al. (1996),
Sakamoto et al. (1999), Sanders et al. (1988), Soifer et al. (1999),
and the NED database.
\label{sed}}
\end{figure}

From the restframe visible to near-IR, the magnification factors are
estimated to be $\sim$~100, and the intrinsic SED, basically
unaffected by differential magnification, is extremely flat ($f_{\nu}
\propto \nu^{-1}$, Figure~\ref{sed}).  On the other hand, in the
restframe mid-IR/far-IR/submm, the exact magnification factors are not
known.  Therefore, we derive instead the possible range of the
intrinsic SED allowed by the model.

The allowed range can be calculated based on the fact that the lens
model requires the magnification factor to be 4--120 in the restframe
mid-IR/far-IR/submm.  In Figure~\ref{sed}, we empirically fit
the observed SED with the following expression,
\begin{equation}
  f_{\nu_{obs}} \propto B_{\nu_{e}}(T_{d}=200 {\rm K}) (1-e^{-\tau_{\nu_{e}}}),
  \label{sed_fit}
\end{equation}
where $\tau_{\nu_{e}}$ is the optical depth parametrized as,
\begin{equation}
  \tau_{\nu_{e}}=\left(\frac{\nu_{e}}{\nu_{0}}\right)^{2}.
  \label{optdep}
\end{equation}
Here, $\nu_{0}$ is set to be 1.5 THz (i.e., 200 $\mu$m).
The possible range of the intrinsic SED (the shaded area in
Figure~\ref{sed}) is determined by demagnifying the observed SED fit
by factors between 4 and 120.  The upper limit corresponds to the case
in which the dust emissivity is low (0.03) and therefore the 200 K dust
region is large, resulting in a small magnification factor while the
lower limit corresponds to the case in which the dust emissivity is
unity (i.e., blackbody) and therefore the 200 K region is small,
resulting in a large magnification factor.

As seen in Figure~\ref{sed}, in principle the intrinsic SED of APM
08279+5255 could strongly peak at $\lambda_{rest} \sim 20 \mu$m.
However, such an SED is unlikely.  If the SED strongly peaks in the
restframe mid-IR, the dust covering factor around the central energy
source must be large.  However, as can be seen in Figure~\ref{sed},
the intrinsic SED is extremely flat at $\lambda_{rest} < 2 \mu$m,
which means that our line of sight to the central QSO is relatively
dust free.  Therefore, if the SED strongly peaks in the mid-IR, it
must mean that we are looking into a heavily dust-enshrouded QSO
through a relatively transparent hole, which seems unlikely.  We
believe it more likely that the intrinsic SED is relatively flat up to
$\lambda_{rest} = 20 \mu$m and drops at longer wavelengths
(dark-shaded region in Figure~\ref{sed}).

\subsubsection{Bolometric Luminosity}
If we assume that the intrinsic SED is flat up to the restframe
mid-IR, the characteristic magnification factor of APM 08279+5255 can
be taken as $\sim 100$.  Therefore, the intrinsic bolometric
luminosity is estimated to be $5 \times 10^{13} {\rm L}_{\odot}$.  The
IR luminosity at $\lambda > 10 \mu$m is $1 \times 10^{13} {\rm
L}_{\odot}$.  Because most of the luminosity is coming out at shorter
wavelengths, we can determine the luminosities without knowing the
precise shape of the restframe far-IR/submm SED.

\subsubsection{Dust Properties}
Figure~\ref{sed} shows that the mid-IR/far-IR/submm SED of APM
08279+5255 is roughly consistent with those of lower-redshift
IR-luminous galaxies/QSOs.  However, since we cannot constrain the
intrinsic SED at $\lambda_{rest} > 100 \mu$m to better than a factor
of 10, it is not possible to set meaningful limits on intrinsic
properties of this object such as the mass and temperature of
dust.

The main sources of uncertainty are the dust emissivity and the
effects of differential magnification.  As seen in
equation~(\ref{obs_flux}), the magnification factor is inversely
proportional to the dust emissivity.  Since the dust emissivity is not
well known even in the local universe, this introduces a large
uncertainty in the magnification factor, and therefore in the
intrinsic luminosity of the dust emission.  In addition, there is a
possibility that the magnification factor may systematically decrease
at longer wavelengths because cooler regions have larger spatial
extent (Eisenhardt 1996; Blain 1999).  If this is the case, there may
exist a cold (e.g., 50 K) massive dust component which is not seen in
the observed spectrum due to its small magnification factor.  Because
of these uncertainties, a self-consistent model can be constructed
with a broad range of dust masses ($10^{6}-10^{9} {\rm M}_{\odot}$)
and temperatures (50--200 K).

Figure~\ref{sed} also illustrates the difficulty of determining dust
properties without knowing the exact shape of the mid-IR/far-IR/submm
SED.  For example, one noticeable feature in the figure is that most
of the objects have comparable submillimeter luminosities in spite of
the large spread in the bolometric luminosity.  This strongly
suggests that the dust distribution in these objects are
``matter-bounded'' (cf. Barvainis et al. 1995): in other words, these
objects are likely to have comparable dust masses, and the difference
in the IR ($>10 \mu$m) luminosities is caused by the difference in dust
temperature (from 50 K (ULIRGs) to 200 K (PG 1206+459) for the dust
component dominating the restframe submm emission).  On the other
hand, the large submillimeter luminosity of BRI1202-0725 indicates
that even if the dust temperature is at 200 K, its dust mass is still
as large as those of local ULIRGs, and could be larger by an order of
magnitude if the dust temperature is low (e.g., 50 K).  The fact that
the allowed IR/submm SED range for APM 08279+5255 is broadly
consistent with the SEDs of such a variety of objects indicates that
with the currently available data, little can be said about the dust
mass and temperature of APM 08279+5255.

\subsection{The Lensing Galaxy}
If the three-image model is correct, there is no observational
image that can be associated with the lensing galaxy.  The redshift of
the lensing galaxy is not determined by the model because the redshift
of the lens enters only through the Einstein radius, and this can be
produced by a distant massive galaxy or a nearby low-mass galaxy.

The derived core radius of 0\farcs2 corresponds to a physical length
of 1.3--1.6 kpc at a redshift of $0.5 < z < 4$.  A galaxy with a core
radius of this size must be a large elliptical-type galaxy with a
velocity dispersion $\sigma_{v}$ of $\sim$ 300 km/s.  To be compatible
with the derived small Einstein radius, the lensing galaxy must be at
a high redshift, probably $z\sim3$, although at the moment there is no
observational evidence to support this hypothesis.  The mass of such a
lensing galaxy is calculated to be $\sim 2 \times 10^{11}$ L$_{\odot}$
within the Einstein radius of 0\farcs29.  Such a galaxy would not have
been detected by any observations to date of this system.

\section{SUMMARY AND CONCLUSIONS}

We have obtained high-resolution images of APM 08279+5255 in the
$K$-band (FWHM $\sim$ 0\farcs15) and at 12.5 $\mu$m (FWHM $\sim$
0\farcs4).  We have constructed a gravitational lens model using the
$K$-band image, and determined the magnification factors at longer
wavelengths based on the 12.5 $\mu$m image and other existing data.
The basic conclusions are as follows:
\begin{enumerate}
  \item APM 08279+5255 consists of three components, which are
  distributed over $\sim$ 0\farcs4.  The third image (component C) is
  bright ($K=14.5$ mag) and compact (FWHM $\sim 0\farcs15$).

  \item Component C could be either a third lensed image (three-image
  model) or the luminous QSO/AGN nucleus of the lensing galaxy
  (two-image model).  If the three-image model is correct, the
  magnification factor is high ($\sim 100$) and the effects of
  differential magnification could be significant.  If the two-image
  model is correct, the magnification factor is low ($\sim 10$), and
  the effects of differential magnification are negligible.  The
  three-image model is preferred by a number of observational
  arguments, and therefore we adopt this model to deduce the
  intrinsic properties.

  \item The derived magnification factors for APM 08279+5255 are 90,
  90--120, and 4-120 in the restframe UV/optical, near-IR, and
  mid-IR/far-IR/submillimeter, respectively.  The corresponding
  intrinsic source radii are $< 20$ pc, $<$ 50 pc, and 60--1800
  pc.  

  \item By assuming that the intrinsic SED is flat up to
  $\lambda_{rest} < 20 \mu$m, we estimate an overall magnification
  factor of $\sim 100$.  From this, the intrinsic bolometric
  luminosity of APM 08279+5255 is derived to be $5\times10^{13} {\rm
  L}_{\odot}$.  Its intrinsic IR luminosity is $1\times10^{13} {\rm
  L}_{\odot}$.  Therefore, APM 08279+5255 is intrinsically luminous,
  but it is not the most luminous object known.

  \item With the currently available data, neither the dust mass nor
  the dust temperature can be determined due to the uncertainties
  associated with the dust emissivity and the possible effects of
  differential magnification.

\end{enumerate}
\acknowledgments

We thank Drs.\ Ibata and Lewis for communicating us their results
prior to the submission of their paper.  E.E.\ thanks Drs.\ Chris
Fassnacht and Andrew Blain for helpful discussions.  The W. M. Keck
Observatory is operated as a scientific partnership between the
California Institute of Technology, the University of California, and
the National Aeronautics and Space Administration.  It was made
possible by the generous financial support of the W. M. Keck
Foundation.  Infrared astronomy at Caltech is supported by grants from
the NSF and NASA.  This research has made use of the NASA/IPAC
Extragalactic Database (NED) which is operated by the Jet Propulsion
Laboratory, California Institute of Technology, under contract with
the National Aeronautics and Space Administration.

\clearpage

\begin{deluxetable}{rccc}
\tablecaption{Newly Measured Flux Densities of APM~08279+5255 \label{new_flux}}
\tablehead{\colhead{$\lambda$} & \colhead{Band} & \colhead{Magnitude}
& \colhead{Flux density} \\[.2ex]
\colhead{($\mu$m)} & \colhead{} & \colhead{(mag)} & \colhead{(mJy)}
}
\startdata
1.25 &  J         & 13.34 $\pm$ 0.03       & 7.19 $\pm$ 0.20 \\
1.65 &  H         & 12.65 $\pm$ 0.03       & 8.97 $\pm$ 0.25 \\
2.15 &  Ks        & 12.08 $\pm$ 0.03       & 9.40 $\pm$ 0.26 \\
3.5  &  L\arcmin\ &  9.90 $\pm$ 0.04       & 27.3 $\pm$ 1.0  \\
12.5 &            &  6.35 $\pm$ 0.15       &   74 $\pm$ 11   \\
17.9 &            &  5.2  $^{+0.8}_{-0.4}$ &  103 $\pm$ 50   \\
\enddata
\end{deluxetable}

\begin{deluxetable}{ccccccccc}
\tablecaption{Positions and brightnesses of three point sources \label{imconfig}}
\tablehead{\colhead{Components} & \colhead{$\Delta$X} & \colhead {$\Delta$Y} & 
\colhead{$\Delta \alpha$} & \colhead{$\Delta \delta$} &
\colhead{Flux ratio} & \colhead{K} & 
\colhead{$\chi^{2}$/dof} & \colhead{FWHM} }
\startdata
 A    &-0\farcs060 & 0\farcs377 & 0\farcs191 &  0\farcs330 & 1.28 & 12.82 & 0.93 & 0\farcs15 \\
 B    & 0\farcs000 & 0\farcs000 & 0\farcs000 &  0\farcs000 & 1.00 & 13.09 & 0.93 & 0\farcs15 \\
 C    &-0\farcs002 & 0\farcs245 & 0\farcs090 &  0\farcs228 & 0.26 & 14.55 & 0.92 & 0\farcs15 \\
Total &            &            &             &             &      & 12.08 &      &           \\
\enddata
\end{deluxetable}

\begin{deluxetable}{lrr}
\tablecaption{The Lens Models \label{bestfit}}
\tablewidth{5in}
\tablehead{\colhead{Parameter} & \colhead{Three-image} & \colhead {Two-image}}
\startdata
Einstein radius ($\theta_{e}$)            & 0\farcs29                  & 0\farcs19  \\
Ellipticity ($\varepsilon$)               & 0.012                      & 0.083 \\
Core radius ($\theta_{c}$)                & 0\farcs21                  & 0\arcsec    \\
Position angle (E of N)                   & 79                         & 100  \\
Lens position\tablenotemark{a}            & (0\farcs074, 0\farcs205)& (0\farcs090, 0\farcs228) \\
Source position\tablenotemark{a}          & (0\farcs069, 0\farcs196)& (0\farcs067, 0\farcs184) \\
Magnification for A \tablenotemark{b}     & 34                         & 3 \\
Magnification for B \tablenotemark{b}     & 43                         & 4 \\
Magnification for C \tablenotemark{b}     &  9                         &   \\
Total Magnification\tablenotemark{b}      & 86                         & 7 \\
\enddata
\tablenotetext{a}{The positions are ($\Delta \alpha$, $\Delta \delta$)
from component B.}
\tablenotetext{b}{The magnification factors are for a point source.}
\end{deluxetable}

\begin{deluxetable}{lccc}
\tablecaption{The Magnification Factors for the Three-Image Model \label{magtable}}
\tablehead{\colhead{} & \colhead{UV/Optical} &
\colhead{Near-IR} & \colhead{Mid-IR/Far-IR/Submm}}
\startdata
Magnification      & 90            & 90--120 & 4--120   \nl
Source radius (pc) & $<$ 20 &  $<$ 50 & 60--1800 \nl
\enddata
\end{deluxetable}

\clearpage

\end{document}